\documentclass[journal=ancac3,manuscript=article,usetitle=true]{achemso}

\usepackage[version=3]{mhchem} 

\author{René Petersen}
\author{Thomas Garm Pedersen}
\email{tgp@nano.aau.dk}
\affiliation{Department of Physics and Nanotechnology, Aalborg University, DK-9220 Aalborg East, Denmark}
\author{Antti-Pekka Jauho}
\affiliation{Department of Micro- and Nanotechnology, Technical University of Denmark, DTU Nanotech, Building 345 East, DK-2800 Kongens Lyngby, Denmark}
\alsoaffiliation{Department of Applied Physics, Aalto University, P. O. Box 11100, FI-00076 AALTO, Finland}

\title{Clar Sextet Analysis of Triangular, Rectangular and Honeycomb Graphene Antidot Lattices}

\begin{document}
\begin{abstract}
Pristine graphene is a semimetal and thus does not have a band gap.
By making a nanometer scale periodic array of holes in the graphene
sheet a band gap may form; the size of the gap is controllable by
adjusting the parameters of the lattice. The hole diameter, hole
geometry, lattice geometry and the separation of the holes are
parameters that all play an important role in determining the size
of the band gap, which, for technological applications, should be at
least of the order of tenths of an eV. We investigate four different
hole configurations: the rectangular, the triangular, the rotated
triangular and the honeycomb lattice. It is found that the lattice
geometry plays a crucial role for size of the band gap: the
triangular arrangement displays always a sizable gap, while for the
other types only particular hole separations lead to a large gap.
This observation is explained using Clar sextet theory, and we find
that a sufficient condition for a large gap is that the number of
sextets exceeds one third of the total number of hexagons in the
unit cell. Furthermore, we investigate non-isosceles triangular
structures to probe the sensitivity of the gap in triangular
lattices to small changes in geometry.
\end{abstract}

{\bf keywords} graphene, antidots, Clar sextets, band structure, band gap


Graphene, a one atom thick layer of carbon, has attracted a great
deal of attention since its discovery in 2004 \cite{ar_geim1}. This
is due to its intriguing properties such as extremely high
conductivity \cite{ar_xudu1}, high mechanical strength \cite{ar_lee}
and the ability to probe relativistic phenomena at sub-light speeds
\cite{ar_castro}. Due to the large conductivity and the atomic layer
thickness, graphene is a promising candidate as a substitute for the
present principal component of most semiconductor devices, silicon.
Natural graphene, however, is a semimetal and thus lacks a band gap
which is a necessary condition for its usage in transistor
architectures \cite{ar_castro}. Introducing a band gap into graphene
can be achieved by various means and several approaches have been
suggested. For example, slicing graphene into graphene nanoribbons
\cite{ar_barone1} or growing graphene epitaxially on a substrate
opens up a band gap in graphene \cite{ar_zhou}.

Recently however, another approach to opening up a gap in graphene
has been suggested. Calculations
\cite{ar_tgp1,ar_rp1,ar_tgp2,ar_tgp4,ar_furst,ar_balog,ar_vanevic} show that
by making a triangular array of holes in the graphene layer a band
gap is obtained and the size of the gap can be tuned by varying the
parameters of the lattice, \textit{i.e.}, the lattice geometry, the hole size
and the hole separation. Several recent theoretical articles have
explored various aspects of graphene antidot lattices, \textit{e.g.},
electron-phonon coupling \cite{ar_vukmir,ar_stojanovic}, detection of edge states
\cite{ar_wimmer}, or details of band gap scaling
\cite{ar_liu,ar_martinazzo}. Graphene antidot lattices have also
been subject to recent experimental research and antidot lattices of
various geometries have been fabricated using a number of different
techniques \cite{ar_balog,ar_eroms,ar_bai,ar_bieri,ar_kim}.

In earlier work triangular antidot lattices have been treated in
detail \cite{ar_tgp1,ar_rp1,ar_tgp2,ar_tgp4,ar_furst,ar_vanevic}, and it was
found that the size of the band gap is directly linked to the size
of the hole compared to the size of the unit cell: the larger the
hole the larger the band gap. To make a thorough analysis, one must
consider other lattice geometries as well in order to assess whether
other geometries might be suited for the actual production of
graphene antidot lattices, and also to determine how sensitive the
lattices are to small structural variations. Indeed, graphene
antidot lattices produced by lithography \cite{ar_eroms} and block co-polymer
masks \cite{ar_bai} will be subject to some uncontrollable variations in the
lattice and thus it is important to examine how large an effect
these variations may have.

\begin{figure}[htb]
    \includegraphics[width=8.5cm]{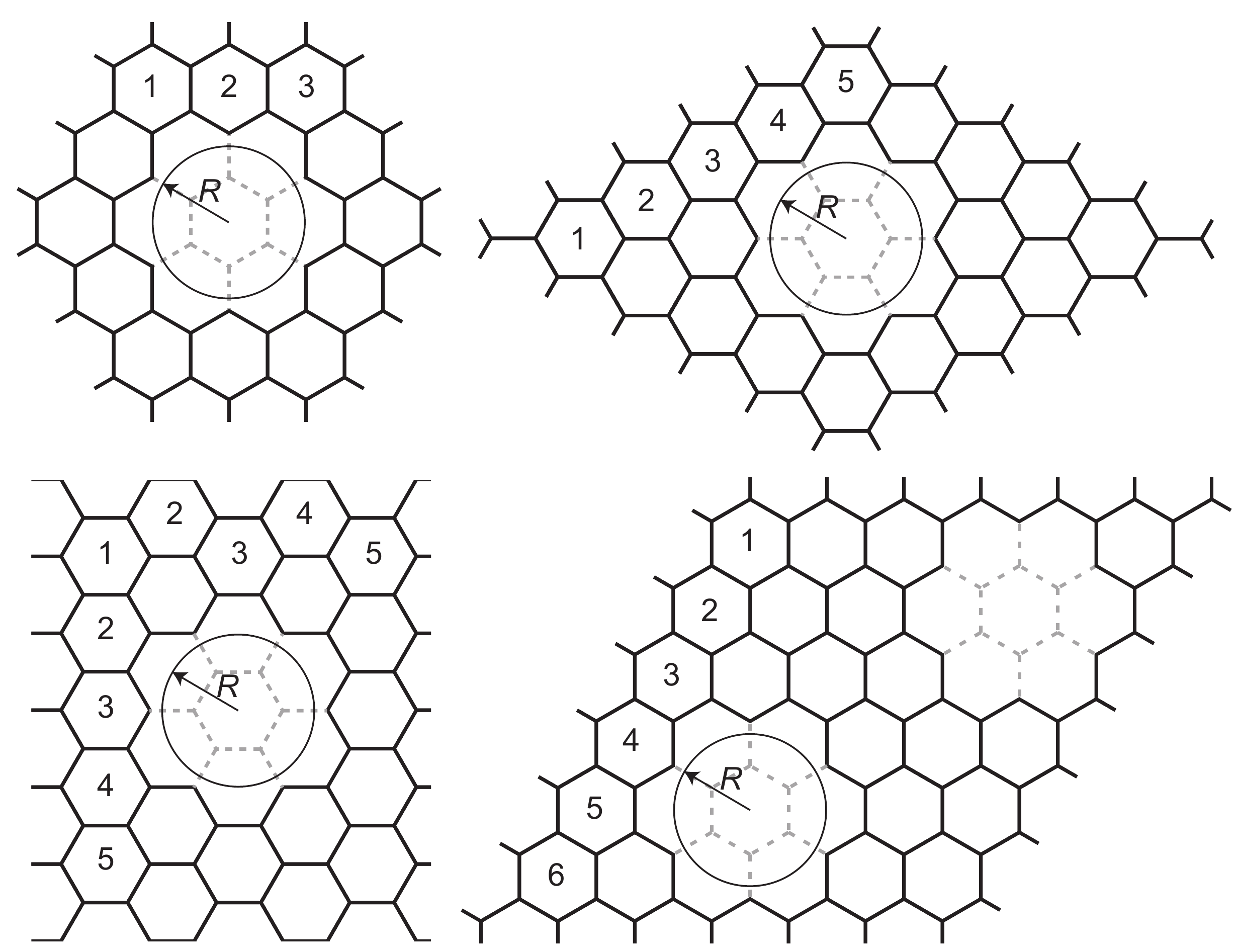}
    \caption{Unit cells of the four types of geometries studied in this paper. Upper left (UL): \{3,1\} triangular lattice, UR: \{5,1\} rotated triangular lattice, LL: \{5,5,1\} rectangular lattice, LR: \{6,1\} honeycomb lattice. Note that the graphene sheet is rotated 90$^\circ$ in the UL and LR illustration. The numbering in each unit cell shows the non-shared hexagons defining the lattices.}
    \label{fig:1}
\end{figure}

This paper is structured as follows.  In Sect. II we introduce the
four different lattices we study, as well as the important concept
of Clar sextets.  Sect. III discusses some computational details,
while in Sect. IV we give our main results, and their analysis.
Finally, a short conclusion is given in Sect. V.

\section{Antidot lattice geometries}
We consider four different lattice types: the triangular lattice,
the rotated triangular lattice, the rectangular lattice and the
honeycomb lattice. In the following, $R$ is always the radius of the
hole given in units of the graphene lattice constant
$a_0=2.46\text{\AA}$.
\paragraph{Triangular.} The holes are oriented in a triangular geometry and the unit cell is denoted as \{$L$, $R$\} where $L$ is the number of non-shared (belonging to only a single unit cell) hexagons on the edge of the unit cell. This is illustrated in the upper left part of \ref{fig:1} where a \{3,1\} unit cell is shown. The numbering in the figure shows the 3 non-shared hexagons. In these geometries, the elementary antidot lattice vectors are parallel to the carbon-carbon bonds.
\paragraph{Rotated triangular.} The holes are oriented as in the triangular geometry but rotated 30$^\circ$. The unit cell is denoted as \{$L$, $R$\} where $L$ is the number of non-shared hexagons on the edge of the unit cell. This is illustrated in the upper right part of \ref{fig:1} where a \{5,1\} unit cell is shown. The elementary antidot lattice vectors are rotated 30$^\circ$ with respect to the carbon-carbon bonds.
\paragraph{Rectangular.} The holes are located on the corners of a rectangle. The unit cell in this geometry is denoted by \{$L_x$, $L_y$, $R$\} where $L_x$ is the number of non-shared horizontal hexagons and $L_y$ the number of non-shared vertical hexagons in the unit cell. Hence, $L_x$ must be odd to keep the unit cell strictly rectangular. This geometry is illustrated in the lower left part of \ref{fig:1} for a \{5,5,1\} lattice.
\paragraph{Honeycomb.} The holes are placed such that they form a honeycomb lattice similar to that of the carbon atoms in a graphene sheet. The unit cell in this geometry is denoted by \{$L$, $R$\} where $L$ is the number of non-shared hexagons on the edge of the unit cell. A \{6,1\} unit cell is shown in the lower right part of \ref{fig:1}. The centre to centre distance between the holes should be $(L+1)a_0/\sqrt{3}$ and the vector between the holes should be at an angle of 30$^\circ$ relative to the zig-zag direction of the graphene sheet for the holes to form a honeycomb lattice. If the first hole is placed such that the centre is exactly in the middle of a hexagon it will not always be such that the centre of the second hole, when placed according to the prescriptions above, is also in the middle of a hexagon. This might cause the holes to be non-similar with respect to the edge of the holes. It turns out that only for unit cells obeying $L=3n+2$ (with $n$ an integer) two similar holes can be placed according to the above prescriptions. For the rest, one of the holes must be displaced slightly to make sure that the centre of both holes is in the middle of a hexagon, thereby ensuring that the two holes are similar. The non-perfect honeycomb lattices differ from the other lattices by their reduced symmetry of the unit cell. Thus, one should be careful when calculating band structures because the irreducible Brillouin zone is larger than for the other geometries.

The selection of structures mentioned above is motivated by recent
experimental work. Honeycomb lattices have been produced by
patterned hydrogen adsorption \cite{ar_balog}, rectangular lattices
have been produced using lithography \cite{ar_eroms}, triangular
lattices have been produced using block copolymer methods
\cite{ar_bai,ar_kim} and rotated triangular structures have been
produced using a method based on surface-assisted coupling of
designed molecular building blocks in Ref. \cite{ar_bieri}. The fact
that "`hypothetical"' structures are studied experimentally
emphasizes the need for theoretical investigations to guide the
experimental work and possibly the fabrication of devices based on
graphene antidot lattices.

To examine the structures we will calculate band structures of the
lattices and analyze their Clar structure, that is, the pattern of
delocalized $\pi$-orbital phenyl ring structures, \textit{i.e.} Clar sextets \cite{bk_clar}.
Clar analysis has previously been used with success to explain the
oscillating behaviour of the band gap in graphene nanoribbons
\cite{ar_baldoni}, and the stability and band gap of carbon
nanotubes \cite{ar_baldoni2}.  Very recently, we gave a preliminary
discussion of lattice-dependence of band gaps in rectangular
graphene antidot lattices \cite{ar_rp2}. The Clar structure of a
given unit cell of a lattice is determined by locating the pattern
of sextets, which gives the maximum number of sextets in the unit
cell. The sextets cannot be distributed freely within the unit cell
due to two limitations: The Clar representation has to preserve the
unit cell (if it failed to do so, it would not, by definition, be a
unit cell) and two sextets cannot be neighbors. Neighboring sextets
are non-chemical since they would require carbon atoms with more
than four bonds. In most cases it is straightforward to determine
the Clar structure while in others it is more involved due to lack
of symmetry. In those cases we have calculated the bond order to aid
in finding the optimal Clar structure. Here it should be noted that
in many cases the Clar structure is not unique. For many structures
several different Clar structures yield the same total number of
sextets. Thus, when calculating the bond order one will find a
superposition of all the distinct Clar structures. This is not
crucial, because, as it will be explained below, what really matters
for our purpose is the number of sextets.

\section{Results and discussion}
\begin{figure}[htb]
    \includegraphics[width=8.5cm]{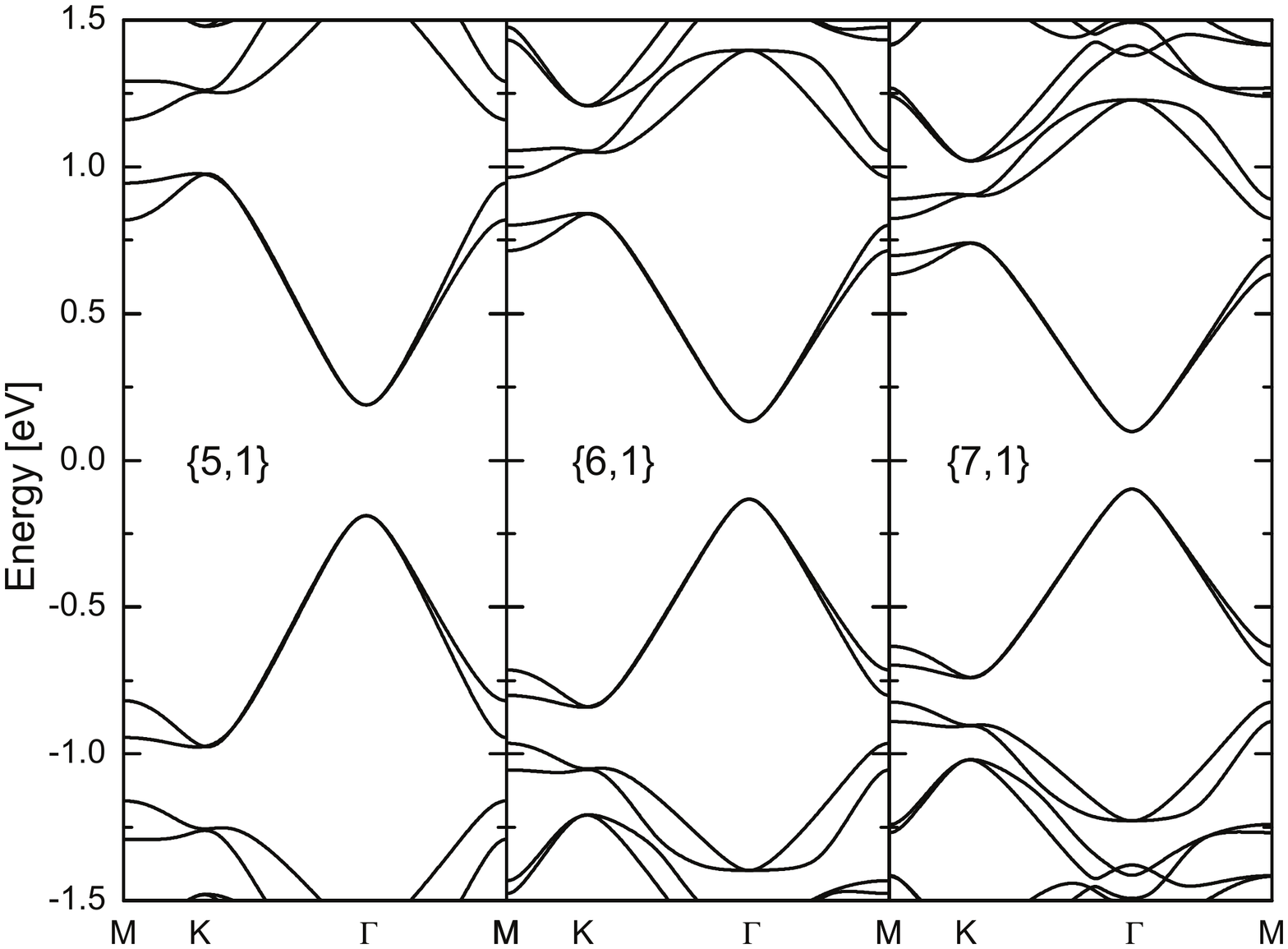}
    \caption{Band structure of the triangular antidot lattices \{5,1\}, \{6,1\} and \{7,1\}. A band gap is present for all structures and it is always located at the $\Gamma$ point of the Brillouin zone.}
    \label{fig:2}
\end{figure}

\begin{figure}[htb]
    \includegraphics[width=8.5cm]{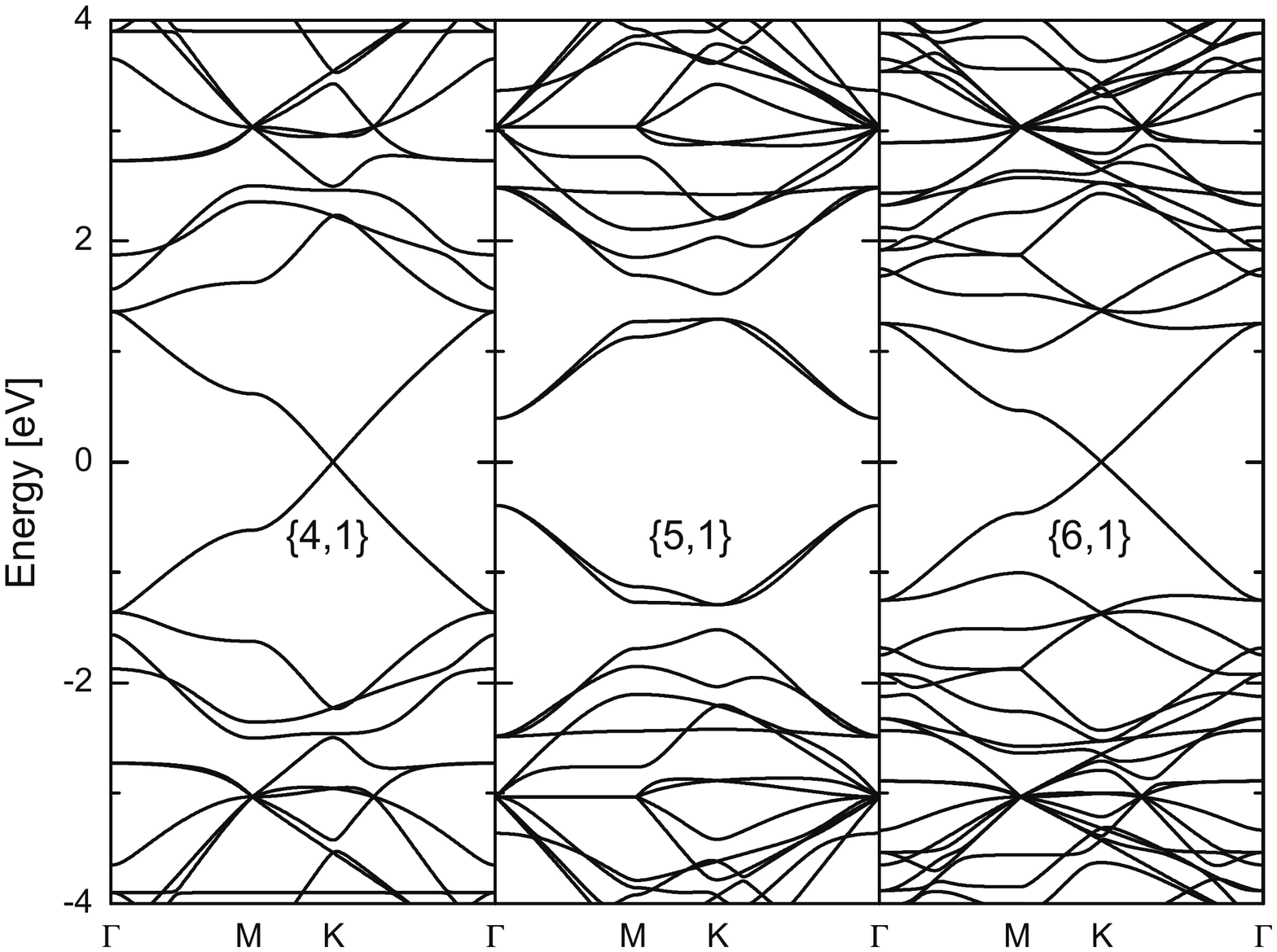}
    \caption{Band structure of the rotated triangular structures \{4,1\}, \{5,1\} and \{6,1\}. Only one of the structures shown possesses a band gap while the others resemble the behaviour of intact graphene near the $K$ point.}
    \label{fig:3}
\end{figure}

\begin{figure}[htb]
    \includegraphics[width=8.5cm]{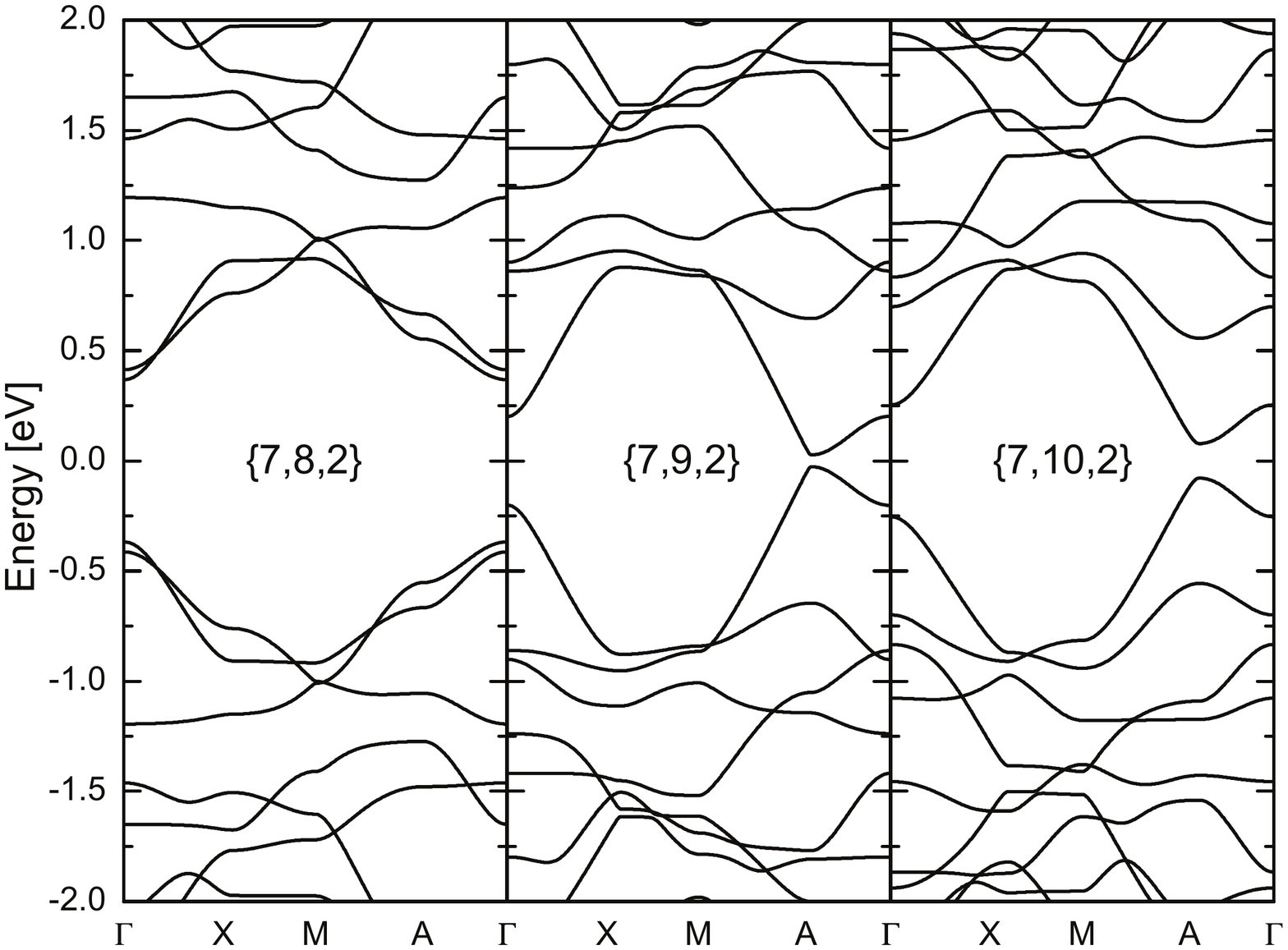}
    \caption{Band structure of the rectangular lattice structures \{7,8,2\}, \{7,9,2\} and \{7,10,2\}. Only one of the structures shown possesses a band gap while the others resemble the behaviour of intact graphene.}
    \label{fig:4}
\end{figure}

\begin{figure}[htb]
    \includegraphics[width=8.5cm]{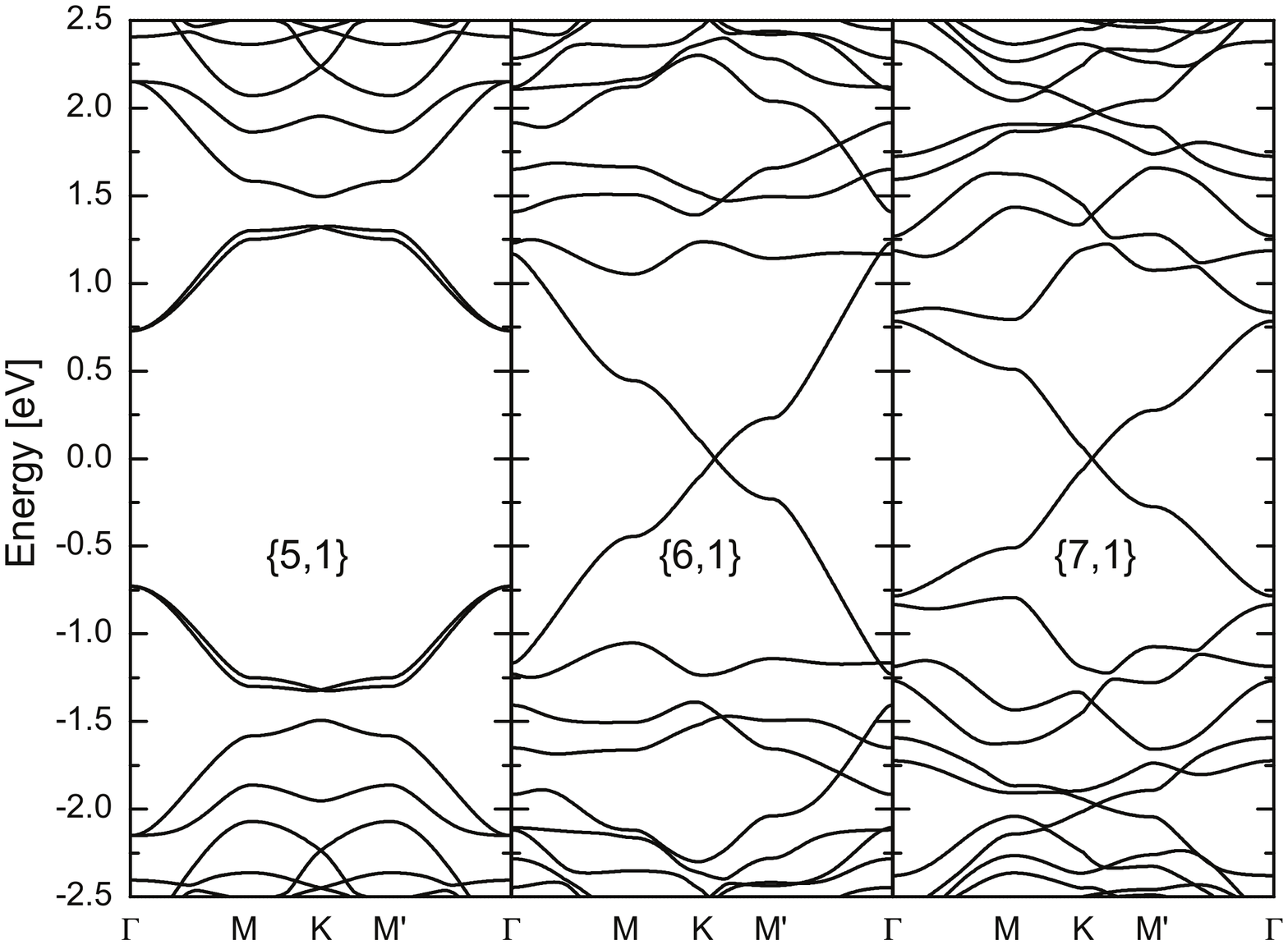}
    \caption{Band structure of the honeycomb lattices \{5,1\}, \{6,1\} and \{7,1\}. A band gap is only present for one of the shown structures.}
    \label{fig:5}
\end{figure}

The results of the band structure calculations of the NN-TB model
are shown in \ref{fig:2}-\ref{fig:5}.
\ref{fig:2} shows the band structure of three triangular antidot
lattices differing in the unit cell size, that is, the separation
between the holes. As shown previously \cite{ar_tgp1,ar_rp1}
triangular antidot lattices show a band gap for all tested
configurations and the band gap $E_g$ is proportional to the ratio
between the number of atoms removed to form the hole and the total
number of atoms in the unit cell before the hole is formed:
$E_g\propto N_{\text{removed}}^{1/2}/N_{\text{total}}$
\cite{ar_tgp1,ar_rp1}. To illustrate the fact that the band gap simply
decreases monotonously with unit cell size for a fixed hole, we have
considered \{5,1\}, \{6,1\} and \{7,1\} triangular lattices. As clearly
observed in \ref{fig:2}, all the
chosen structures have large band gaps and the band gap is always
located at the $\Gamma$ point of the Brillouin zone. Indeed, it is
observed that the band gap decreases as the ratio between the hole
size and unit cell size decreases, that is, as the ratio
$N_{\text{removed}}^{1/2}/N_{\text{total}}$ decreases.

The story is different for other geometries, as demonstrated by
\ref{fig:3}, which gives the band structures for the rotated
triangular lattices. No band gap is observed for the structures
\{4,1\} and \{6,1\} and around the $K$ point the bands resemble the
bands of pristine graphene: no gap is observed and the bands are
linear in the proximity of the $K$ point. As we shall discuss below,
for general structures of the type \{$L$,$R$\} only every third value of $L$
leads to a substantial band gap.

When analyzing the band structures of rectangular and
honeycomb/near-honeycomb (because the lattice is disrupted to make
the holes similar) antidot lattices the picture is similar to the
rotated triangular lattices. For the three rectangular lattices
shown in \ref{fig:4} all structures have a finite band gap but
only \{7,8,2\} presents a large band gap, while \{7,9,2\} and
\{7,10,2\} have significantly smaller band gaps. For the three
honeycomb lattices in \ref{fig:5} only \{5,1\} presents a large
gap. These findings strongly suggest that some  connection should
exist between certain general characteristics of the lattice and the
appearance of a large band gap. It should be pointed out that the
band gap is not exactly zero for any of the shown structures but it
is indeed very small in magnitude (on the order of few meV).

\begin{figure}[htb]
    \includegraphics[width=8.5cm]{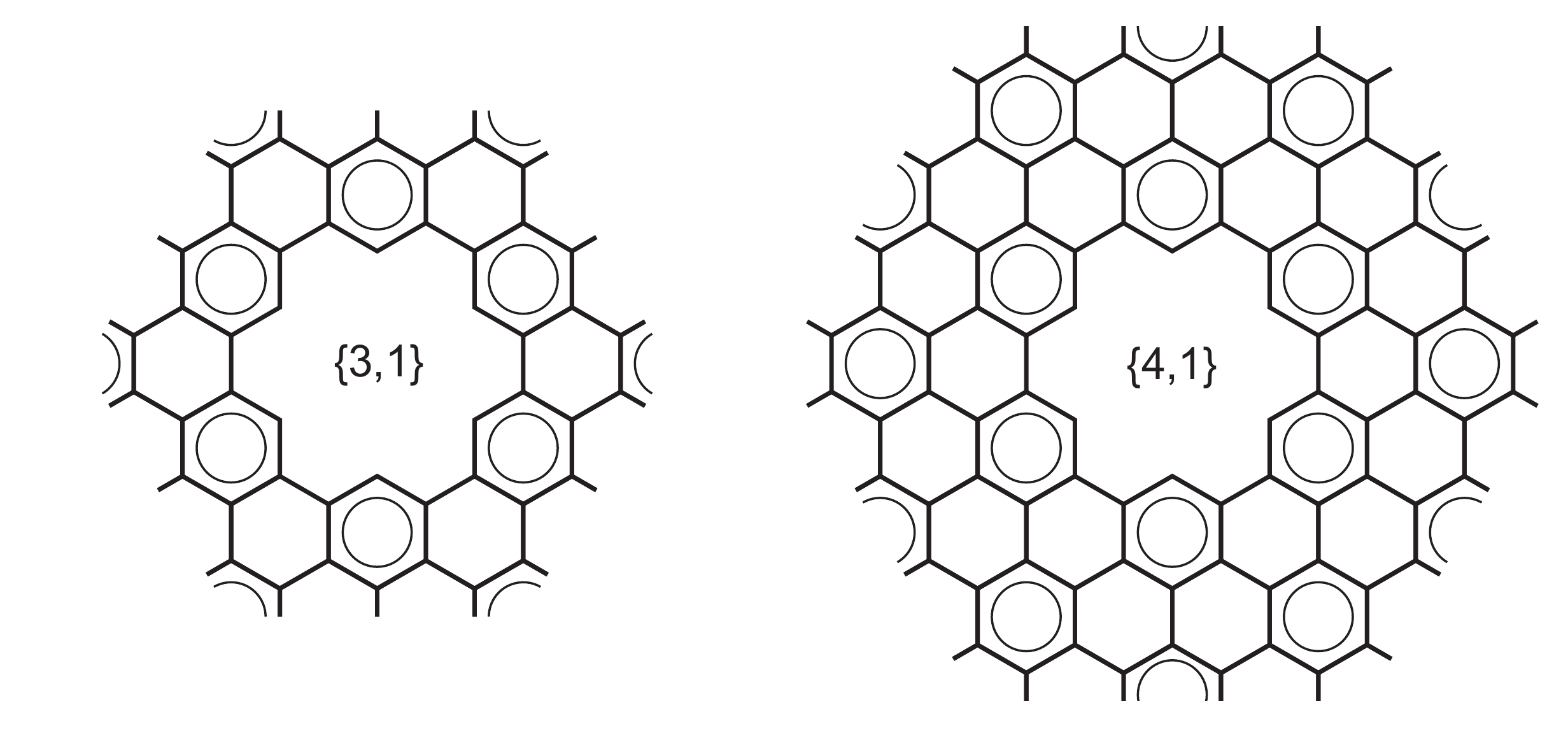}
    \caption{Clar structure of the triangular lattice. Here it is clear that all lattices support a complete benzenoid structure for a hole of radius $a_0$.}
    \label{fig:6}
\end{figure}

To explain the presence of a large band gap for certain structures
and the lack of a band gap for other structures we suggest that one
should analyze the Clar representation of the unit cell. By doing
this one finds that not all of the structures support a complete
benzenoid pattern because the Clar sextets cannot be distributed
freely across the unit cell. The Clar representation of the
triangular lattice is particularly simple because it always allows
for a complete benzenoid structure just like in pristine graphene.
In other words, the introduction of the holes does not disturb the
structure of the resonant double bonds and thus the resonant
structure remains the same as in pristine graphene. The only
exception to this rule is related to the double bonds around the
hole which, depending on the radius of the hole, may not be allowed
to maintain their chemical structure. \ref{fig:6} shows the Clar
structure of two unit cells belonging to the triangular lattice,
\{3,1\} and \{4,1\}. Both structures support the complete benzenoid
pattern. According to Clar sextet theory \cite{ar_baldoni} fully
benzenoid structures have higher stability than structures for which
a fully benzenoid bonding pattern is not possible. Thus, one can
expect triangular lattices to be more stable than other geometries.

\begin{figure}[htb]
    \includegraphics[width=8.5cm]{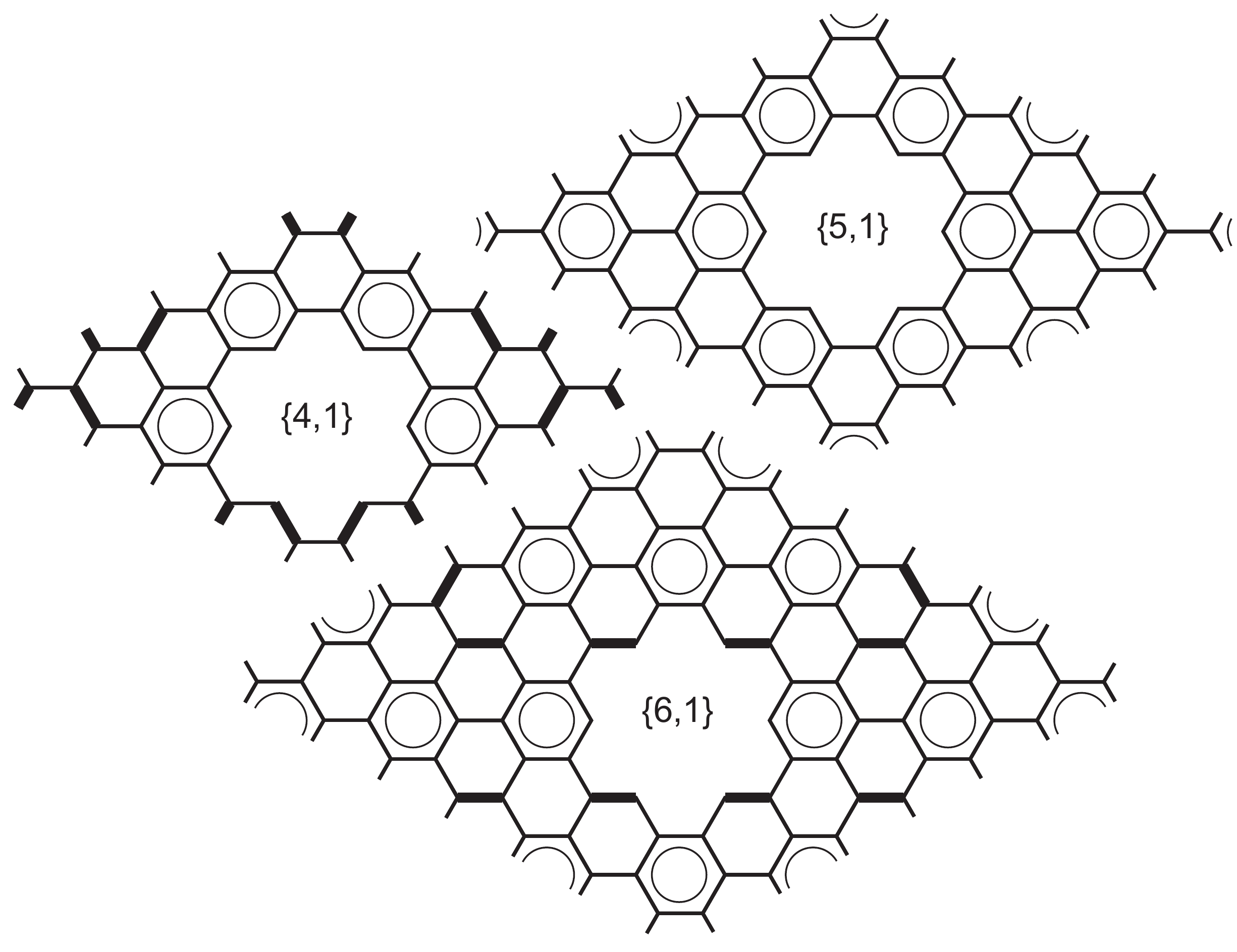}
    \caption{Clar structure of the triangular rotated lattice. As it is seen, a complete benzenoid pattern is not possible for all structures. The hole radius is $a_0$ in all cases.}
    \label{fig:7}
\end{figure}

\begin{figure}[htb]
    \includegraphics[width=8.5cm]{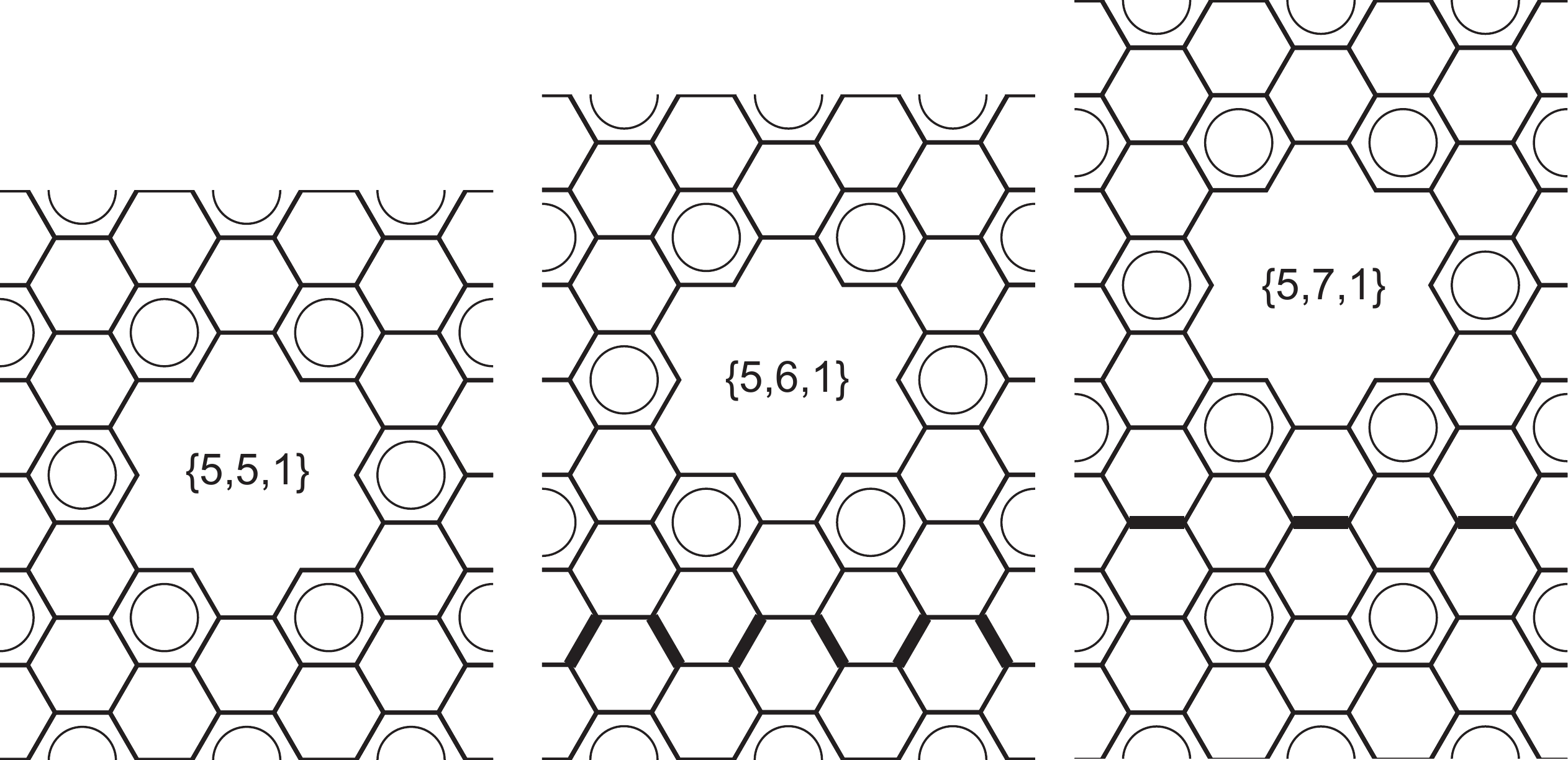}
    \caption{Clar structure of the rectangular lattice. As it is seen, a complete benzenoid pattern is not possible for all structures. The hole radius is $a_0$ in all cases.}
    \label{fig:8}
\end{figure}

\begin{figure}[htb]
    \includegraphics[width=8.5cm]{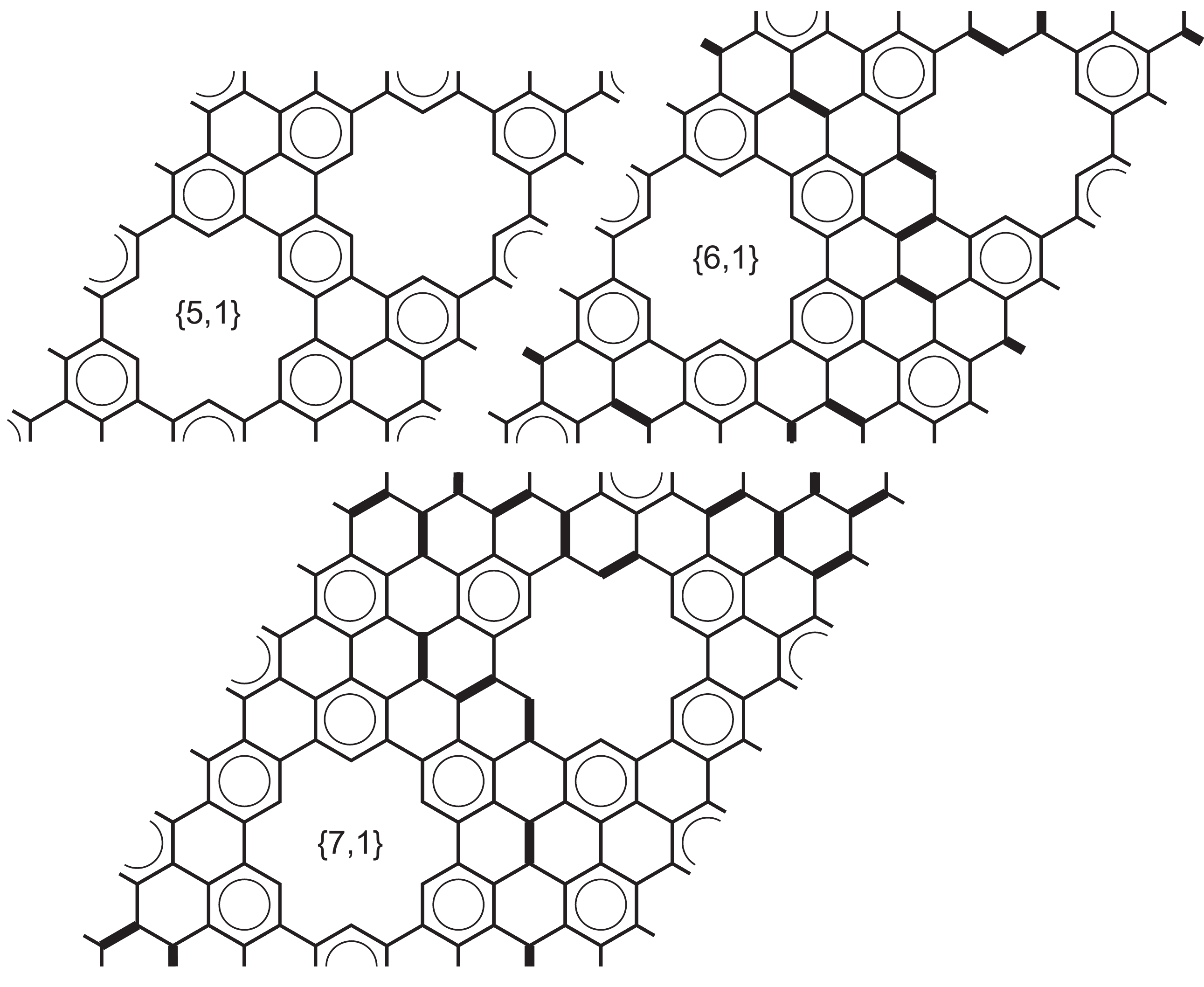}
    \caption{Clar structure of the honeycomb lattice. As it is seen, a complete benzenoid pattern is not possible for all structures. The hole radius is $a_0$ in all cases.}
    \label{fig:9}
\end{figure}

\begin{table}[htb]
    \centering
    \begin{tabular}{|c|c|}
        \hline
        Structure & large band gap \\
        \hline
triangular & no restrictions \\
rotated triangular & $L=3n+2 $\\
rectangular & $L_y=3n+2$\\
honeycomb & $L=3n+2$\\
 \hline
    \end{tabular}
    \caption{Empirical rules governing the occurrence of a large band gap. $n$ is a non-negative integer.}
    \end{table}

For the other lattice types a complete benzenoid pattern is not
always a possibility. This becomes evident by studying \ref{fig:7}-\ref{fig:9}. From these figures one can also see that
the 3-periodic patterns found for the band gaps are replicated in
the Clar patterns of the structures. Thus, only those structures,
which have a fully benzenoid pattern lead to a large band gap while
the other structures either present a gap that is significantly
smaller (for the rectangular lattices a reduction of a factor 5 is
seen) or practically zero. By combining the calculations of the band
structures  with the Clar representations of the unit cells we may
deduce a set of semi-empirical rules for the occurrence of a large
band gap; these rules are summarized in the Table 1. Thus, the structures with significant band gaps
constitute only one third of the total number of structures within these last three classes of lattices. These findings are based on the NN-TB model but
we have replicated the same patterns in the QT-TB model in order to
verify that the conclusions drawn are not based on artifacts of an
oversimplified model.

\begin{figure}[htb]
    \includegraphics[width=8.5cm]{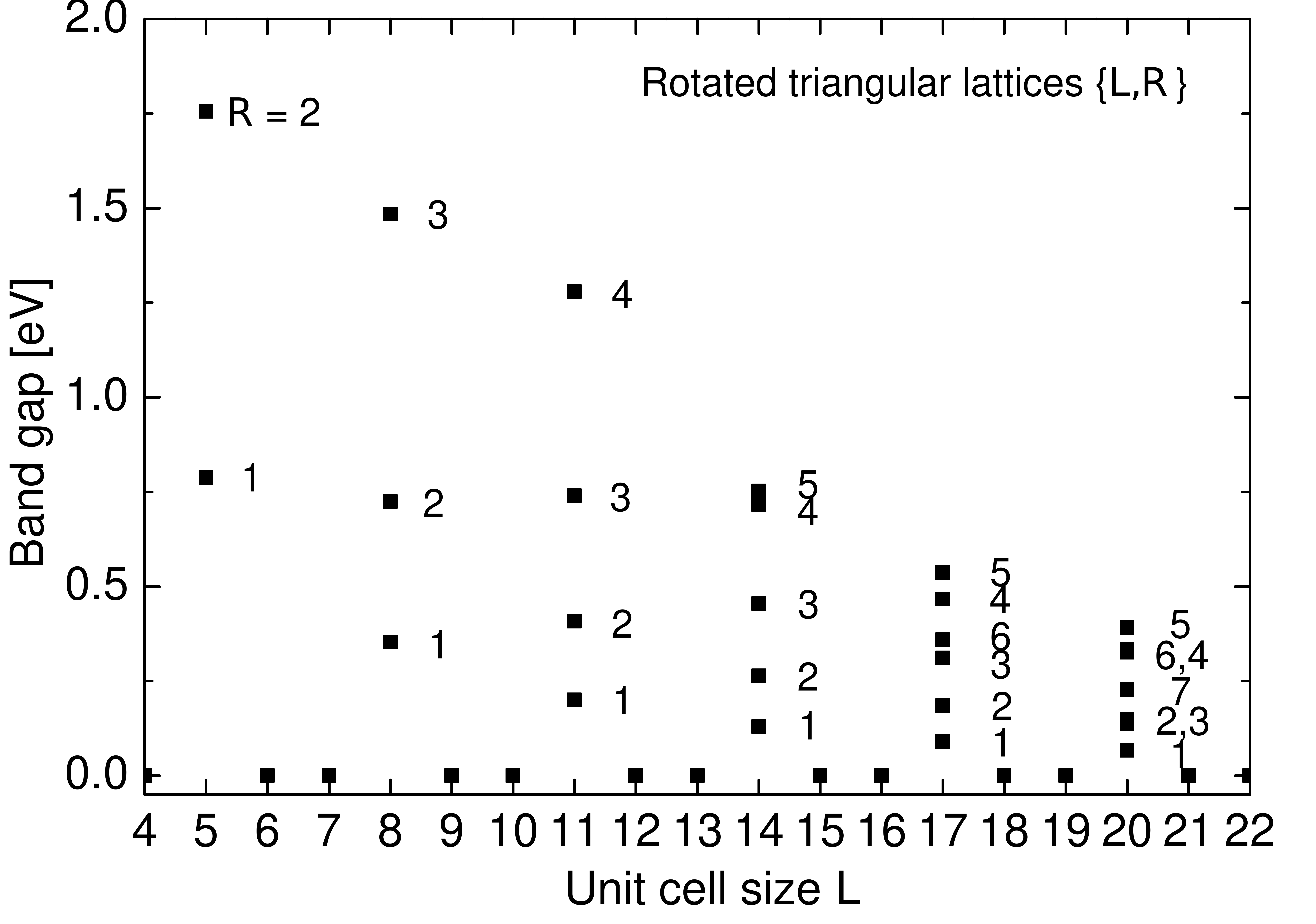}
    \caption{Band gaps of $\{L,R\}$ rotated triangular antidot lattices. The $R$ values are indicated next to each data point.}
    \label{fig:10}
\end{figure}

The possibility of making a complete benzenoid pattern of Clar
sextets in graphene antidot lattices, ignoring the disruption of the
Clar structure by the hole, seems to be a criterion for the
appearance of a large band gap. Evidently, all triangular antidot
lattices do possess a band gap and they all support a complete
benzenoid pattern of sextets. On the contrary, for the other
lattices, only a minority of all structures support a complete
benzenoid pattern and consequently possess a large band gap. For
hole sizes of $R=1$ it seems that the band gap is either large or
close to zero. In order to extend our conclusions to larger (and more realistic) geometries,
we have tested a large number of rotated triangular lattices as a function of the hole size $R=1..7$ and the lattice
spacing $L=5..20$, see \ref{fig:10}. We always find that the prediction of the Clar sextet theory holds:
as a function of $L$ only every third structure has a sizable gap.  The size of the
gap always decreases as $L$ is increased; however the quantitative details depend on the value of $R$, and
are thus beyond the qualitative statements that can be deduced from the Clar theory. Moreover,
in Ref. \cite{ar_rp2} we have verified Clar theory for $R=2$ rectangular lattices.
In general, our calculations indicate
that a criterion for a large band gap is the existence of a complete
benzenoid Clar pattern. In an attempt to find a simple rule for the
existence of a large band gap we counted the number of sextets in
the unit cells and related it to the total number of hexagons in the
cell. We found that, for those structures having a large band gap,
the number of sextets in the unit cell was larger than one third of
the total number of hexagons in the unit cell,
$N_{\text{Sx}}>\tfrac{1}{3}N_{\text{Hx}}$.

\begin{figure}[htb]
    \includegraphics[width=8.5cm]{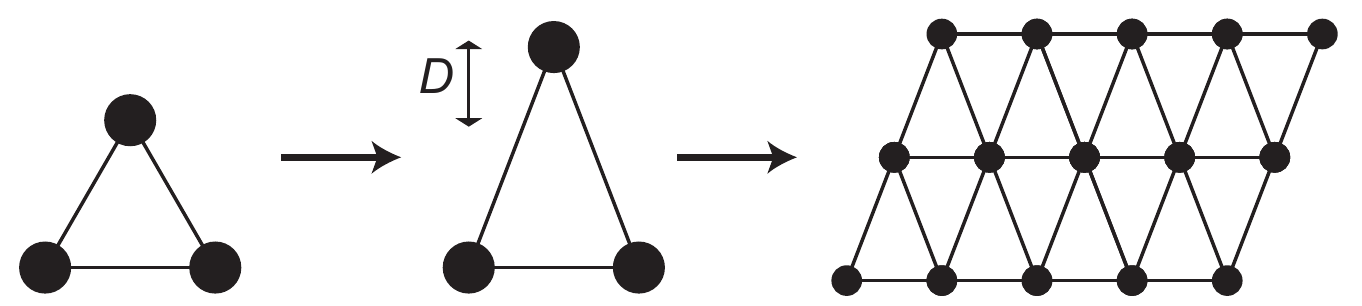}
    \caption{A non-isosceles triangular lattice. This lattice is denoted by \{$L$,$D$,$R$\} where $L$ and $R$ are the same as in the triangular case and $D$ measure the deviation of the hole shown from its original position in the triangular lattice. $D$ is measured in units of $a_0$ and can be negative.}
    \label{fig:11}
\end{figure}

From these findings we conclude that the non-rotated triangular
lattice holds the most potential for the actual production of
graphene antidot lattice of technological importance, since a band
gap is found in all cases. Thus, it is interesting to study the
stability of this structure under small geometric distortions. Here
we will consider a non-isosceles triangular lattice as shown in
\ref{fig:11}. All unit cells are effectively elongated in the
$y$-direction and the deviation from the triangular case is denoted
with the parameter $D$, which expresses how much the vertical
distance between the holes is larger than in the triangular case.
$D$ is measured in units of $a_0$ and a lattice of the non-isosceles
triangular type is denoted as \{$L$, $D$, $R$\} where $L$ retains
its original meaning. The elongation of the unit cell in the
$y$-direction disturbs the previously complete benzenoid structure
and one could suspect that a similar three fold repetitive pattern
as those seen for other types of structures should be seen. Indeed,
if one analyzes the Clar pattern it is found that every third
structure, those with $D=3n$, support a complete benzenoid Clar
structure. Looking a \ref{tbl:1} one can see that these
structures are exactly those, which also possess a band gap, in
accordance with the findings for other structures.

Thus, as a guide to experimental fabrication of large-gap antidot
lattices we stress the following points: First, triangular lattices
are favorable due to their insensitivity to the precise lattice
constant. It is essential, however, that the elementary lattice
vectors connecting neighboring holes are aligned along the
carbon-carbon bonds. Also, it is important to maintain six-fold
rotational symmetry as demonstrated by the analysis of non-isosceles
lattices. In practice, orientation of the antidot lattice relative
to the graphene lattice requires knowledge of the latter. This may
be obtained by electron diffraction \cite{ar_knox}, transmission
electron microscopy \cite{ar_meyer} or polarized Raman spectroscopy \cite{ar_huang}.
Controlling the orientation of the lattice should be feasible
with lithography \cite{ar_eroms} but probably challenging with the
block copolymer technique \cite{ar_bai}. Alternatively, chemical
self-assembly from suitable precursors can be applied to ensure a
particular lattice geometry \cite{ar_bieri}.

\begin{table}[htb]
    \centering
    \begin{tabular}{|c|c|c|c|}
        \hline
        Structure & $E_g \text{[eV]}$ & Pos $\vec{b_1}$ & Pos $\vec{b_2}$ \\
        \hline
\{5,0,1.0\}$^*$  & 0.377  & 0.000 & 0.000 \\ \hline
\{5,1,1.0\}      & 0.044  & 0.667 & 0.333 \\ \hline
\{5,2,1.0\}      & 0.052  & 0.667 & 0.333 \\ \hline
\{5,3,1.0\}$^*$  & 0.269  & 0.000 & 0.000 \\ \hline
\{5,4,1.0\}      & 0.033  & 0.667 & 0.333 \\ \hline
\{5,5,1.0\}        & 0.089  & 0.667 & 0.333 \\ \hline
\{5,6,1.0\}$^*$  & 0.210  & 0.000 & 0.000 \\ \hline
    \end{tabular}
    \caption{Band gaps and position of band gaps within the Brillouin zone of non-isosceles triangular structures. Structures which allow for complete benzenoid structures are marked with a star. \label{tbl:1}}
\end{table}

\section{Conclusion}
Our results  show that it is possible, without turning to full-scale
atomistic calculations, predict if a given graphene antidot
structure can be expected to possess a large band gap only by
analyzing the Clar structure of the unit cell. Structures
investigated in this work show a large band gap only if the lattice
allows for a complete benzenoid pattern with the number of sextets
exceeding one third of the total number of hexagons in the unit
cell. Four different lattice types were investigated. We found that
only non-rotated triangular lattices, in which antidot lattice
vectors are parallel to atomic bonds, are insensitive to lattice
constants and always exhibit a band gap. All other lattices (rotated
triangular, rectangular and honeycomb) are extremely sensitive to
the lattice geometry and only one third display large band gaps.
Finally, non-isosceles triangular lattices show the same three-fold
repetitive pattern with respect to the band gap.

\section{Methods}
In the present work, band structures of antidot lattices are calculated in a simple
nearest neighbor tight binding model (NN-TB) as well as the quasi-particle tight binding (QP-TB)
model \cite{ar_rp1} based on the parametrization of the
quasi-particle band structure of graphene \cite{ar_gruneis1}. In the NN-TB
model the hopping integral between neighbor atoms is given by
$\gamma=3.033\text{eV}$ \cite{bk_saito} and overlap is neglected. In
the QP-TB model the parameters are used as given in Ref.
\cite{ar_gruneis1} and three nearest neighbors and overlaps are
included in the calculations.

In certain cases, the Clar structure is difficult to identify and for this
purpose the bond order (BO) pattern has been examined. In graphene and related structures one
can calculate the BO between two bound atoms by calculating
the overlap between the $\pi$-electrons of the two atoms. This gives
information about the probability of finding a double bond between
those two atoms. The BO between atom $p$ and $p'$ (neighboring
atoms) is calculated as follows

\begin{align}
BO_{pp'}=\sum_{v}\sum_{v'} \left({c_v^{p}}\right)^*c_{v'}^{p'}.
\end{align}

Here, $c_v^p$ is the expansion coefficient of valence band state $v$
in the basis of $\pi$-orbitals labeled by their site $p$, and the
sums are taken over all valence band states $v,v'$. In the present
case, this entails a summation over $k$-points in the irreducible
Brillouin zone as well as band index. A large BO is indicative of
double-bond character and the BO pattern is therefore helpful in identifying
the Clar pattern. We do not explicitly show the obtained BO patterns but merely
ensure their agreement with all presented Clar structures.

\acknowledgement
Financial support from the Danish FTP Research council grant "Nano
engineered graphene devices" is gratefully acknowledged.  A.-P. Jauho is
grateful to the FiDiPro program of Academy of Finland.

\providecommand*\mcitethebibliography{\thebibliography}
\csname @ifundefined\endcsname{endmcitethebibliography}
  {\let\endmcitethebibliography\endthebibliography}{}

\end{document}